\documentclass{emulateapj}
\shorttitle{SN~2010jl Progenitor}
\shortauthors{Smith et al.}
\begin{document}

\title{A Massive Progenitor of the Luminous Type~II\lowercase{n}
  Supernova 2010\lowercase{jl}\altaffilmark{1}}

\author{Nathan Smith\altaffilmark{2}, Weidong Li\altaffilmark{3}, Adam
  A.\ Miller\altaffilmark{3}, Jeffrey M.\ Silverman\altaffilmark{3},
  Alexei V.\ Filippenko\altaffilmark{3}, Jean-Charles
  Cuillandre\altaffilmark{4}, Michael C.\ Cooper\altaffilmark{5},
  Thomas Matheson\altaffilmark{6}, \& Schuyler D.\ Van
  Dyk\altaffilmark{7}}

\altaffiltext{1}{Based in part on observations made with the NASA/ESA
  Hubble Space Telescope, obtained at the Space Telescope Science
  Institute, which is operated by the Association of Universities for
  Research in Astronomy, Inc., under NASA contract NAS5-26555.}
\altaffiltext{2}{Steward Observatory, University of Arizona, 933 North
  Cherry Avenue, Tucson, AZ 85721; Email:
  nathans@as.arizona.edu.} 
\altaffiltext{3}{Department of Astronomy, University of California,
  Berkeley, CA 94720-3411.}
\altaffiltext{4}{Canada-France-Hawaii Telescope Corporation, 65-1238
  Mamalahoa Hwy., Kamuela, HI 96743.}
\altaffiltext{5}{Department of Physics and Astronomy, University of
  California, 4129 Frederick Reines Hall, Irvine, CA 92697-4575.}
\altaffiltext{6}{National Optical Astronomy Observatory, 950 North
  Cherry Avenue, Tucson, AZ 85719-4933.} 
\altaffiltext{7}{Spitzer Science Center, California Institute of
  Technology, Mail Code 220-6, 1200 East California Blvd., Pasadena,
  CA 91125.}

\begin{abstract}

  The bright, nearby, recently discovered supernova (SN)~2010jl is a
  member of the rare class of relatively luminous Type~IIn events.
  Here we report archival {\it Hubble Space Telescope} ({\it HST})
  observations of its host galaxy UGC~5189A taken roughly 10~yr prior
  to explosion, as well as early-time optical spectra of the SN.  The
  {\it HST} images reveal a bright, blue point source at the position
  of the SN, with an absolute magnitude of $-$12.0 in the F300W
  filter.  If it is not just a chance alignment, the source at the SN
  position could be (1) a massive young ($<$6 Myr) star cluster in
  which the SN resided, (2) a quiescent, luminous blue star with an
  apparent temperature around 14,000~K, (3) a star caught during a
  bright outburst akin to those of luminous blue variables (LBVs), or
  (4) a combination of option 1 and options 2 or 3.  Although we
  cannot confidently choose between these possibilities with the
  present data, any of them imply that the progenitor of SN~2010jl had
  an initial mass above 30~M$_{\odot}$.  This reinforces mounting
  evidence that many SNe~IIn result from very massive stars, that
  massive stars can produce visible SNe without collapsing quietly to
  black holes, and that massive stars can retain their H envelopes
  until shortly before explosion.  Standard stellar evolution models
  fail to account for these observed properties.

\end{abstract}

\keywords{circumstellar matter --- stars: evolution --- stars: mass
  loss --- stars: winds, outflows --- supernovae: general}

%%%%%%%%%%%%%%%%%%%%%%%%%%%%%%%%%%%%%%%%%%%%%%%%%%%%%%%%%%%%%%%%%%%%%%%%%%
\section{INTRODUCTION}

Supernova (SN)~2010jl was discovered on 2010 Nov.\ 3.52 (UT dates are
used throughout this paper) by Newton \& Pucket (2010).  With a
discovery magnitude of 13.5 (unfiltered), this is one of the brightest
supernovae (SNe) in recent years.  After one day it continued to
brighten (12.9 mag on 2010 Nov. 4.50), signaling that this SN was also
caught early in its evolution.  Moreover, its host galaxy UGC~5189A is
located at a distance of almost 50 Mpc, suggesting that SN~2010jl is
intrinsically very luminous, with an absolute magnitude of about $-$20
at a time when it was still becoming brighter.  Early-time spectra
showed that it is a Type~IIn SN (Benetti et al.\ 2010; see Filippenko
1997 for a review of SN types). Although SNe~IIn constitute about
6--9\% of core-collapse SNe (Smith et al.\ 2010c; Li et al.\ 2010),
SN~2010jl appears to be a member of the class of unusually luminous
examples of these (Smith et al.\ 2007, 2008, 2010a; Prieto et
al.\ 2007; Drake et al.\ 2010; Rest et al.\ 2009).

In this paper we analyze the pre-explosion archival images of the
field of SN~2010jl obtained with the {\it Hubble Space Telescope}
({\it HST}). We have obtained ground-based post-explosion images of
the SN that allow us to constrain its position. We find a blue source
in the {\it HST} images that is coincident with the SN position to
within 1$\sigma$ of our astrometric solution, suggesting that the
source is likely to be either a detection of the blue progenitor star
itself, or the star cluster in which it resided (or both).  As
discussed below, this progenitor candidate has important implications
for SNe~IIn, as well as for the evolution and death of massive stars
in general.

%%%%%%%%%%%%%%%%%%%%%%%%% FIGURE 1 - finder charts %%%%%%%%%%%%%%%%%%%%%%%%%
\begin{figure*}
\epsscale{0.95}
%\plotone{prog.eps}
\plotone{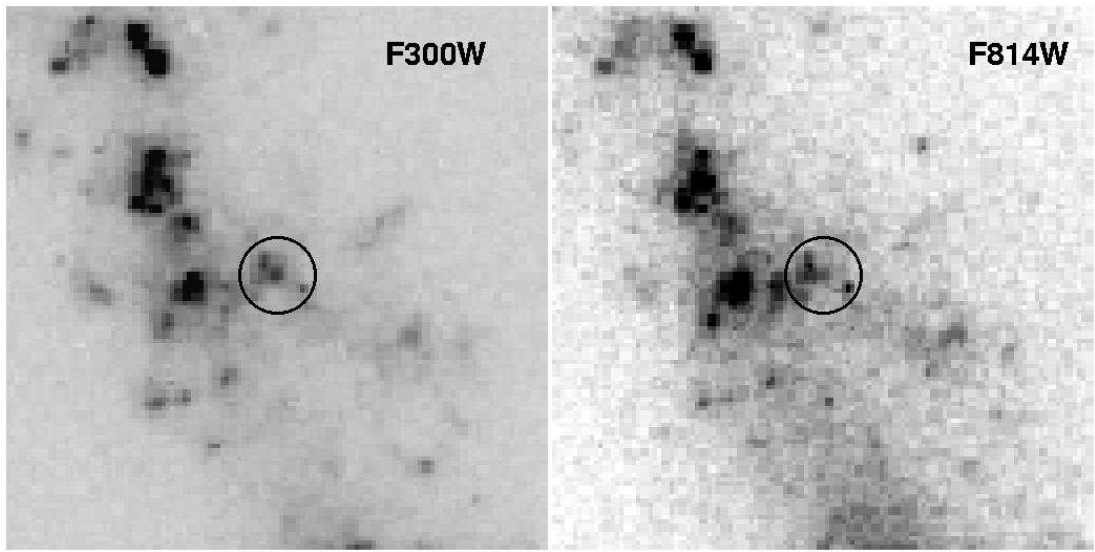}
\caption{Images of the environment ($10\arcsec \times 10\arcsec$) of
  SN 2010jl in its host galaxy, UGC~5189A. These are $HST$/WFPC2
  images in the (a) F300W and (b) F814W filters, obtained in 2001
  Feb.\ North is up and east is to the left. The circle has a radius
  of 0$\arcsec$.47, which is 10 times the 1$\sigma$ uncertainty of our
  astrometric solution.}
\label{fig:finder}
\end{figure*}
%%%%%%%%%%%%%%%%%%%%%%%%%%%%%%%%%%%%%%%%%%%%%%%%%%%%%%%%%%%%%%%%%%%%%%

There have been two previous detections of progenitors of SNe~IIn, and
both were luminous stars that reinforce a suspected link between
SNe~IIn and the class of massive unstable stars known a luminous blue
variables (LBVs).  One case is SN~2005gl, which was a moderately
luminous SN~IIn that transitioned into a more normal SN~II.
Pre-explosion images showed a source at the SN position that
disappeared after the SN faded (Gal-Yam \& Leonard 2009).  The high
luminosity suggested that the progenitor was a massive LBV (Gal-Yam \&
Leonard 2009).  The other example of a claimed detection of a SN~IIn
progenitor --- SN~1961V --- has a more circuitous and complicated
history. For decades SN~1961V was considered a prototype (although the
most extreme example) of giant eruptions of LBVs, and an analog of the
19th century eruption of $\eta$~Carinae (Goodrich et al.\ 1989;
Filippenko et al.\ 1995; Van Dyk et al.\ 2002).  However, two recent
studies (Smith et al.\ 2010b; Kochanek et al.\ 2010) argue for
different reasons that SN~1961V was probably a true core-collapse SN,
or at least that it shares observed properties with SN generally
considered to be core-collapse SNe IIn.  Both studies point out that
the pre-1961 photometry of this source's variability may be both a
detection of a very luminous quiescent star, as well as a precursor
LBV-like giant eruption in the few years before core collapse.
SN~1961V is, however, a complicated case and study of it continues.

This strong connection between SNe~IIn and LBVs based on their
progenitor stars supports an existing link based on the physics of
SN~IIn explosions --- namely, accounting for highly luminous SNe~IIn
with a blast wave hitting a massive opaque shell (e.g., Smith \&
McCray 2007; van Marle et al.\ 2009) requires strong eruptive mass
loss in the years preceding core collapse, consistent with giant
eruptions of LBVs (Smith et al.\ 2007, 2008, 2010a; Gal-Yam \& Leonard
2009).  There are additional reasons to suspect a connection between
LBVs and SNe~IIn, and these are reviewed elsewhere (Smith 2008).

In this paper we present the third detection of a candidate progenitor
of a SN~IIn.  This adds to a number of candidate detections of other
SN progenitors, most of which are SNe~II-P (summarized by Smartt
2009). Recently, there have also been some claimed detections of
SN~II-L progenitors which suggest progenitor stars that were somewhat
more massive than those of SNe~II-P (Elias-Rosa et al.\ 2010a, 2010b;
see also Leonard 2010).  If true, the more massive progenitors of
SNe~II-L and IIn would require substantial modification to current
views of SN progenitors and massive-star evolution in general (Smith
et al.\ 2010c).

%%%%%%%%%%%%%%%%%%%%%%%%%%%%%%%%%%%%%%%%%%%%%%%%
\section{OBSERVATIONS}

The host galaxy of SN 2010jl had observations taken $\sim$10 yr prior
to discovery with the Wide Field Planetary Camera 2 ({\it HST}/WFPC2),
which we retrieved from the {\it HST} archive and analyzed.  UGC~5189A
was observed in the F300W and F814W filters on 2001 Feb.\ 14 as part
of GO-8645, with exposure times of 1800~s and 200~s, respectively.

To pinpoint the precise location of the progenitor in the {\it HST}
images, we obtained ground-based images of SN~2010jl for comparison
using MegaCam on the 3.6-m Canada France Hawaii Telesope (CFHT).
These images yielded an image quality of 0$\farcs$6 with 0$\farcs$187
pixels.  To perform astrometric solutions between the ground-based and
{\it HST} images, we adopted the technique detailed by Li et al.\
(2007) using stars present in both images. Geometrical transformation
between a combined 600 s $r$-band image (with multiple short 10~s
exposures to ensure that SN 2010jl was not saturated) taken with
MegaCam on 2010 Nov.\ 9.60 and the 2001 {\it HST}/WFPC2 images yields
a precision of 0.47 WFPC2 pixels (0$\farcs$047) for the SN location in
the WFPC2 images. (Note that an independent astrometric solution by
one of us [S.D.V.] finds a larger 1$\sigma$ precision of 0$\farcs$09.)
Within the uncertainty of the SN position, an object is clearly
detected in the F300W image, and marginally detected in the F814W
images, with a position of $\alpha = 9^{\rm h}42^{\rm m}53\fs 33$,
$\delta = +09\arcdeg 29\arcmin 42\farcs06$ (J2000.0).

Figure~\ref{fig:finder} shows a $10\arcsec \times 10\arcsec$ region of
the site of SN~2010jl in the F300W and F814W {\it HST}/WFPC2 images.
A candidate progenitor source is detected within 1$\sigma$ precision
of the astrometric solution.  The {\it HST} photometry for the
progenitor candidate as measured with {\tt HSTphot} (Dolphin 2000a,
2000b) yields F300W = $21.6 \pm 0.06$ mag and F814W = $23.1 \pm 0.18$
mag.  The candidate is surrounded by some faint extended emission and
has a neighboring source within $<$0$\farcs$4, so we forced HSTphot to
recognize the position of the candidate in order to extract the
photometry.  Due to the complicated background, we suspect that the
uncertainties of the photometry from {\tt HSTphot} are significantly
underestimated, especially for the F814W filter image.  The candidate
source itself has a full width at half-maximum intensity (FWHM) less
than 0$\farcs$3, corresponding to $\sim$73 pc at the distance of
UGC~5189A.

%%%%%%%%%%%%%%%%%%%%%%%%% FIGURE 2 - spectrum %%%%%%%%%%%%%%%%%%%%%%%%%
\begin{figure*}
\epsscale{0.95}
%\plotone{../SPECTRA/plotspec.eps}
\plotone{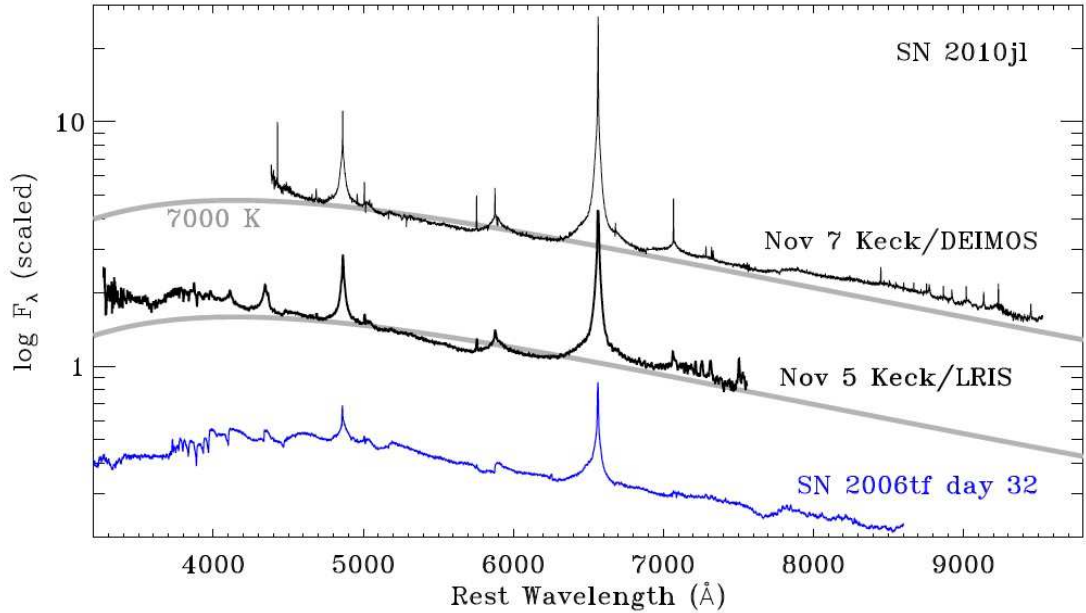}
\caption{Optical spectra of SN~2010jl obtained at early times on 2010
  Nov.\ 5 and 7 (black) compared to the day 32 spectrum of SN~2006tf
  from Smith et al.\ (2008).  All spectra are dereddened by
  $E(B-V)$=0.027 mag (by coincidence, SN~2006tf has the same estimated
  Galactic reddening value; see Smith et al.\ 2008).  A 7000~K
  blackbody is shown in gray for comparison with the SN~2010jl spectra
  (note that a single blackbody component cannot fit the observed
  continuum shape corrected only for Galactic extinction).}
\label{fig:spec}
\end{figure*}
%%%%%%%%%%%%%%%%%%%%%%%%%%%%%%%%%%%%%%%%%%%%%%%%%%%%%%%%%%%%%%%%%%%%%%

We have also initiated a campaign to obtain intensive spectroscopy of
SN~2010jl.  These spectra will be analyzed in detail in a future
paper, but here we briefly discuss the appearance of the early-time
spectrum and the profile of H$\alpha$. Figure~\ref{fig:spec} shows two
spectra of SN~2010jl obtained on 2010 Nov.\ 5 with the Low Resolution
Imaging Spectrometer (Oke et al. 1995) on the 10~m Keck-1 telescope,
and on 2010 Nov.\ 7 with the Deep Imaging Multi-Object Spectrograph
(DEIMOS; Faber et al. 2003) mounted on the 10~m Keck-2 telescope.  All
observations were obtained with the slit oriented at the parallactic
angle (Filippenko 1982).  Standard routines were used to extract and
calibrate the spectra (e.g., Foley et al. 2003).

Figure~\ref{fig:spec} compares our spectra of SN 2010jl with the
early-time (day 32) spectrum of the very luminous SN~IIn 2006tf from
Smith et al.\ (2008).  Although the spectra of SNe~2010jl and 2006tf
are not identical, the continuum shape, Balmer-line strengths and
profiles, and presence of weak He~{\sc i} and other narrow lines are
sufficient to claim that the spectrum of SN~2010jl is consistent with
those of previously observed luminous SNe~IIn.  (There is considerable
variety in the spectra of SNe~IIn.  For a comparison of several other
examples, see Smith et al.\ 2010a, as well as Filippenko 1997.)  The
DEIMOS spectrum, which has significantly higher resolution than the
LRIS spectrum, shows a number of narrow emission and absorption
components from the dense pre-shock circumstellar medium (CSM), which
will be analyzed in more detail in a forthcoming paper (the narrow
H$\alpha$ absorption component is discussed below).
Figure~\ref{fig:spec} also illustrates a 7000~K blackbody for
comparison, which is not a fit.  The mismatch between the 7000~K
blackbody and the observed continuum shape suggests that either
multiple-temperature components are present, or that there is
substantial additional local reddening (and, hence, a higher implied
continuum temperature).

\begin{figure}
\epsscale{0.98}
%\plotone{../SPECTRA/halpha.eps}
\plotone{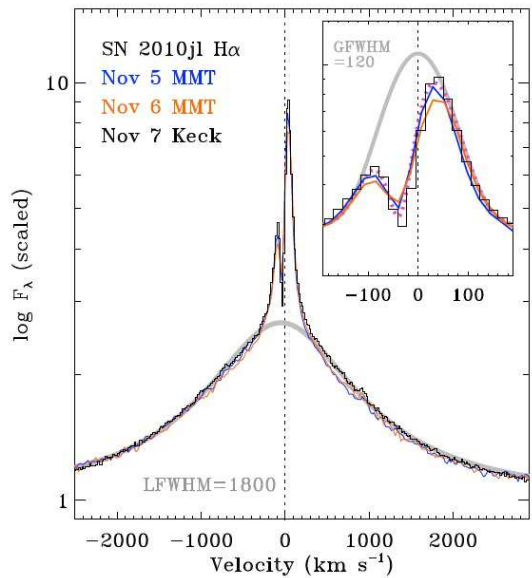}
\caption{The H$\alpha$ profile of SN~2010jl on 2010
  Nov.\ 5 (blue), 6 (orange), and 7 (black histogram), taken with the
  MMT Blue Channel spectrograph and Keck/DEIMOS. These correspond to
  days 2, 3, and 4 after discovery, respectively, and show little
  change with time or observing parameters (see text).  The thick gray
  curve is a Lorentzian profile with FWHM = 1800 km s$^{-1}$.  The inset
  shows the narrow profile on an expanded velocity scale.  The gray
  curve is a symmetric Gaussian with FWHM = 120 km s$^{-1}$, while the
  dotted magenta curve is the same, but with a narrower blueshifted
  Gaussian subtracted (centered at $-$28 km s$^{-1}$, FWHM = 64 km
  s$^{-1}$).}
\label{fig:specH}
\end{figure}

Figure~\ref{fig:specH} shows the high-resolution H$\alpha$ profile of
SN~2010jl observed on 2010 Nov.\ 5, 6, and 7, assuming a redshift $z$
of 0.011.  Spectra on the first two nights were obtained using the
Blue Channel spectrograph mounted on the Multiple Mirror Telescope
(MMT), with 105~s exposures, the 1200 lines mm$^{-1}$ grating, and a
1$\farcs$0 slit width.  The Nov.\ 7 spectrum was obtained with
Keck/DEIMOS, using a resolution of 4400 and a 1$\farcs$0 slit.  The
resulting normalized spectra in Figure~\ref{fig:specH} are remarkably
consistent on all three nights, despite different facilities, setups,
and observing conditions.  This offers reassurance that the
double-peaked narrow profile is not a subtraction artifact that might
arise from oversubtracting a nearby H~{\sc ii} region along the slit,
for example.

%%%%%%%%%%%%%%%%%%%%%%%%% FIGURE 3 - spectrum %%%%%%%%%%%%%%%%%%%%%%%%%
%%%%%%%%%%%%%%%%%%%%%%%%%%%%%%%%%%%%%%%%%%%%%%%%%%%%%%%%%%%%%%%%%%%%%%

The H$\alpha$ profile has an intermediate-width component that can be
approximated by a Lorentzian profile with FWHM = 1800 km s$^{-1}$ (the
thick gray curve in Figure~\ref{fig:specH}), which may be common in
SNe~IIn at early times because of high optical depths (see Smith et
al.\ 2010a).  The wings of this Lorentzian extend to more than
$\pm$4000 km s$^{-1}$. This Lorentzian profile is shifted by $-$50 km
s$^{-1}$, and the high signal-to-noise ratio spectra show some minor
deviations from perfect symmetry in the line wings.

The narrow H$\alpha$ component appears double peaked, and can be
approximated by a symmetric Gaussian emission component with FWHM =
120 km s$^{-1}$ (solid gray curve), but with an absorption component
at $-$28 km s$^{-1}$ relative to the emission-component centroid.  The
$-$28 km s$^{-1}$ absorption suggests that the pre-shock CSM along our
line of sight is rather slow, comparable to the wind speed of an
extreme red supergiant (RSG) that might be a plausible progenitor of a
SN~IIn (Smith et al.\ 2009).  However, the 120 km s$^{-1}$ emission
component suggests that the ionized pre-shock wind in directions away
from our line of sight is faster.  These higher speeds in emission are
faster than what one normally attributes to RSGs, perhaps supporting
the possibility that the progenitor was in an LBV-like phase.
Alternatively, at such early times (and relatively small radii in the
CSM), radiative acceleration of the pre-shock CSM by the SN light may
also play a role (e.g., Chugai et al.\ 2002), although in this case it
would be unclear why the 28 km s$^{-1}$ component along our
line of sight is not accelerated as well. The faster speeds seen in
emission may also indicate an asymmetric pre-shock CSM; we plan to
investigate this asymmetry further in a later paper.

\section{LIKELY INTERPRETATIONS}

We adopt a distance to UGC~5189A of $48.9 \pm 3.4$ Mpc (distance
modulus $m-M = 33.45 \pm 0.15$ mag) and a Galactic reddening value of
$E(B-V) = 0.027$ mag ($A_U = 0.149$ mag, $A_I = 0.053$ mag) from
Schlegel et al.\ (1998).  We do not assume any local host-galaxy
reddening in our analysis below.  With these parameters, the apparent
magnitudes imply a very luminous source with absolute magnitudes of
about $-$12.0 (F300W) and $-$10.4 (F814W).  Possible interpretations
for this luminous blue source are discussed below.

\begin{figure}
\epsscale{0.96}
%\plotone{hstphot.eps}
\plotone{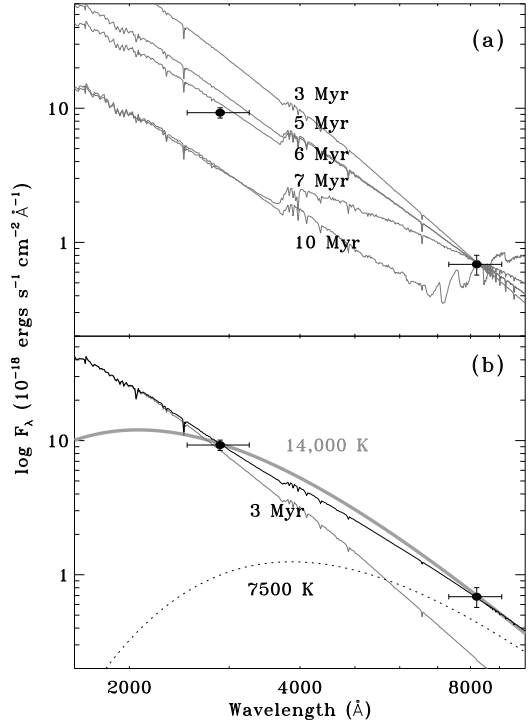}
\caption{The black points in both panels are the fluxes of the
  candidate progenitor derived from the F300W and F814W WFPC2 images,
  dereddened by $E(B-V) = 0.027$ mag as described in the text.  Panel (a)
  compares this photometry to Starburst99 (Leitherer et al.\ 1999)
  models of the integrated spectrum of a massive star cluster with
  ages of 3, 5, 6, 7, and 10 Myr.  Panel (b) shows the same
  photometry, but compared to a 14,000~K blackbody (thick gray line),
  and a composite spectrum (thin black line) that results from the
  combination of a 3~Myr cluster (same as above) and a 7500~K
  blackbody (dotted line), as might be expected from a cool LBV.}
\label{fig:SED}
\end{figure}

1.  {\it The SN progenitor resided in a blue star cluster.}  If the
blue source detected in the {\it HST} image is not dominated by
emission from the progenitor star itself, it could be a luminous blue
star cluster at the same position, of which the progenitor may have
been a member.  Figure~\ref{fig:SED}a shows that the blue color of the
source could be explained by a young star cluster with an age of 5--6
Myr.  If the progenitor candidate of SN~2010jl is actually a young
blue star cluster, it is among the most massive young star clusters
known.  Even in colliding starburst galaxies like the Antennae,
clusters with $M_V$ $<$ --10 mag are extremely rare (Whitmore et al.\
2010).  As a more familiar example in a dwarf irregular galaxy, the
entire 30~Doradus complex has an absolute visual magnitude of about
$-$11, but this would be spread over $\sim$1$\farcs$5 at the distance
of UGC~5189A.  The more compact star cluster R136 in the core of
30~Dor has an absolute magnitude of only about $-$9.3 mag, and would
be spatially unresolved in the {\it HST} images of UGC~5189A.  It is
probable that any member of such a young star cluster reaching core
collapse would be among the most massive stars in that cluster, and a
cluster age of $<$7 Myr implies a stellar lifetime corresponding to
initial masses of $>$30 M$_{\odot}$ (e.g., Schaller et al.\ 1992), if
the cluster is roughly coeval to within about 1~Myr.

%%%%%%%%%%%%%%%%%%%%%%%%% FIGURE 4 - SED %%%%%%%%%%%%%%%%%%%%%%%%%
%%%%%%%%%%%%%%%%%%%%%%%%%%%%%%%%%%%%%%%%%%%%%%%%%%%%%%%%%%%%%%%%%%%%%%

2.  {\it The SN progenitor was an extremely luminous LBV-like star in
  quiescence.} If an absolute F814W magnitude of $-$10.4 corresponds
to an individual star, that star was extremely luminous and massive.
The most massive main-sequence O-type stars do not have visual
luminosities this high, because they are too hot and they emit most of
their flux in the ultraviolet.  To be this bright at red wavelengths,
a star would need to be evolved, shifting its bolometric flux to
longer wavelengths.  However, even the most massive yellow hypergiants
and RSGs have bolometric luminosities fainter than about $-$9.5 mag
(Humphreys et al.\ 1979), and these sources are redder than the
progenitor candidate anyway, so these cannot account for the detected
source.  The dereddened color is consistent with an apparent
temperature of roughly 14,000~K (Figure~\ref{fig:SED}b).  The only
viable type of quiescent blue star would be an extremely luminous
LBV-like star, but it would need a luminosity comparable to the most
luminous known stars such as $\eta$~Car, implying an initial mass
above 80 M$_{\odot}$.

3.  {\it The SN progenitor was normally fainter, but was caught in a
  precursor LBV-like eruption phase.}  One could relax the requirement
that the progenitor of SN~2010jl was among the most massive stars
known if the star was in an outburst state at the time it was observed
by {\it HST}.  An absolute F814W magnitude of $-$10.4 with a blue
color is within the range of observed values for LBV-like eruptions,
either as a bright S~Doradus eruption or a relatively modest example
of a giant LBV eruption (see Smith et al.\ 2010b for details). The
blue color, though, would be more consistent with the latter (Smith et
al.\ 2010b).  This explanation has the advantage that a precursor
LBV-like eruption is needed anyway, in order to create the dense CSM
needed to explain the Type~IIn spectrum and high luminosity of the SN
(e.g., Smith et al.\ 2008).  Since these outbursts can, in some cases,
last for $\sim$10 yr (see Smith et al.\ 2010b), it is not necessarily
improbable to catch a progenitor star in this phase within the decade
before core collapse.

4.  {\it A combination of the above.}  It is also possible that the
detected flux from the progenitor candidate has contributions from
both a host cluster and options 2 or 3 above.  However, as shown in
Figure~\ref{fig:SED}b, if most of the red flux comes from a cool LBV
with an apparent temperature around 7,500~K, for example, then this
tightens the restrictions on the cluster age: the cluster must be
bluer and therefore younger than for a cluster alone, implying an age
of 3~Myr or less. By the same line of reasoning discussed above, this
younger age would imply an even more massive progenitor.

Hypothetically, there is a very small possibility that any of these
three types of sources could be seen at the SN position due to a
chance line-of-sight projection.  For a $20 \times 20$ pixel area
around the SN location ($2\farcs0 \times 2\farcs0$), 13 sources were
detected in the F300W image at the 3$\sigma$ level. For an error
radius of 0.47 pixel, the chance coincidence is only 2.3\%. A chance
projection is therefore very unlikely, and moreover, this type of
ambiguity plagues all studies of SN progenitors and SN host sites.  To
confirm that our candidate source detected in archival data was in
fact the direct detection of a luminous progenitor star will require
additional observations after the SN has faded, to see if it has
significantly changed --- but for a luminous SN~IIn that may continue
to interact with dense CSM, we may need to wait several years.  Even
before that time, however, one significant point is clearly evident:
{\it all three plausible scenarios require the progenitor of
  SN~2010jl to have been a very massive star}, with an initial mass
higher than those typically derived for SNe~II-P (Smartt 2009; Leonard
2010).  This has significant implications for stellar evolution,
discussed next.

\section{IMPLICATIONS FOR MASSIVE-STAR EVOLUTION}

Whether the progenitor candidate is a young star cluster or a direct
detection of the progenitor star itself, the luminous blue source
implies that the progenitor had an initial mass above 30 M$_{\odot}$.
SN progenitors below this range, as seen for SNe II-P (Smartt 2009),
are not found to reside in very luminous, compact, young star clusters.
An individual star with a quiescent luminosity of the candidate
progenitor would have an initial mass $\ga$80 M$_{\odot}$, and a
star caught in an LBV outburst would most likely be a star with an
initial mass above 30 M$_{\odot}$ as well (see Smith et al.\ 2010b).

%One consequence of the progenitor being a massive star is that this SN
%is not one of the type of hybrid Type Ia/IIn explosions that are
%thought to result from a SN~Ia exploding in a dense H-rich CSM, such
%as SN~2002ic or SN~2005gj (Hamuy et al.\ 2003; Aldering et al.\ 2006;
%Prieto et al.\ 2007).  Benetti et al.\ (2006) have questioned whether
%these SN are really Type Ia, or if they may be core-collapse SNe
%instead.  Moreover,

A massive star progenitor for SN~2010jl adds to mounting evidence for
three general conclusions concerning the fates of massive stars:

(1) SNe~IIn arise preferentially from very massive stellar
progenitors.  As noted in \S~1, this is based on the direct detections
of LBV-like progenitors of SN~2005gl and SN~1961V, as well as on the
large amounts of mass in the CSM needed to explain luminous SNe~IIn.
If the SN~2010jl progenitor candidate is a luminous individual star
resembling an LBV, it further strengthens the connection between LBVs
and SNe~IIn.

(2) Since SN~2010jl is a SN~IIn, requiring that the progenitor ejected
H-rich material shortly before core collapse, its massive progenitor
reinforces the conclusion that very massive stars sometimes retain H
envelopes until shortly before core collapse, instead of shedding all
of their H envelopes at nearly solar metallicity to produce SNe~Ibc
(e.g., Heger et al.\ 2003).  A viable alternative, which is consistent
with the observed fractions of various SN subtypes, is that many
SNe~Ibc result instead from close binary evolution across a wide range
of progenitor mass, and that the most massive single stars produce
SNe~IIn (Smith et al.\ 2010c; Yoon, Woosley, \& Langer 2010).

(3) Lastly, if the SN~2010jl progenitor was a massive star, it
provides another example suggesting that very massive stars can
produce luminous explosions, instead of collapsing quietly to a black
hole (see O'Connor \& Ott 2010; Smith et al.\ 2010c).

Standard stellar evolution models fail to account for all three of
these basic observational indications.  Instead, models generally
infer that single massive stars at roughly solar metallicity with
initial masses above 30 M$_{\odot}$ either will shed their H enveloeps
to make SNe~Ibc, or will fail to make successful visible explosions
and collapse directly to black holes (e.g., Heger et al.\ 2003).

%\medskip
\acknowledgments 
\footnotesize
%\scriptsize

%We thank an anonymous referee for critical comments that helped
%improve the manuscript.  

Based in part on observations obtained at the MMT Observatory, a joint
facility of the Smithsonian Institution and the University of
Arizona. Based in part on observations obtained with
MegaPrime/MegaCam, a joint project of CFHT and CEA/DAPNIA, at the
Canada-France-Hawaii Telescope (CFHT) which is operated by the
National Research Council (NRC) of Canada, the Institut National des
Science de l'Univers of the Centre National de la Recherche
Scientifique (CNRS) of France, and the University of Hawaii.  Some of
the data presented herein were obtained at the W.M.\ Keck Observatory,
which is operated as a scientific partnership among the California
Institute of Technology, the University of California, and NASA; the
observatory was made possible by the generous financial support of the
W. M. Keck Foundation. We thank the staffs at these observatories for
their efficient assistance, as well as R.J.\ Foley and S.B.\ Cenko for
their help at Keck.

The supernova research of A.V.F.'s group at U.C. Berkeley is supported
by National Science Foundation grant AST-0908886, by the TABASGO
Foundation, and by NASA through grants AR-11248 and AR-12126 from the
Space Telescope Science Institute, which is operated by Associated
Universities for Research in Astronomy, Inc., under NASA contract NAS
5-26555. J.M.S.\ is grateful to Marc J.\ Staley for a Graduate
Fellowship. We thank J.R.\ Graham and S.\ Wright for obtaining Keck
adaptive optics images of the site of SN 2010jl for us; unfortunately,
these data did not yield a useful astrometric solution because only
the SN was clearly detected.

%KAIT and its ongoing operation were made possible by donations from
%Sun Microsystems, Inc., the Hewlett-Packard Company, AutoScope
%Corporation, Lick Observatory, the NSF, the University of California,
%the Sylvia \& Jim Katzman Foundation, and the TABASGO Foundation.

{\it Facilities:} {\it HST} (WFPC2), Keck I (LRIS), Keck II (DEIMOS),
MMT (Blue Channel), CFHT (MagaCam)

% REFERENCES

\end{document}